\documentclass[twocolumn,showpacs,preprintnumbers,
amsmath,amssymb,aps,floatfix,pra,superscriptaddress]{revtex4}
\usepackage[dvips]{graphicx}
\usepackage{amsfonts}
\usepackage{dcolumn}
\usepackage{bm}
\usepackage{color}
\usepackage{epsfig}
\usepackage{hyperref}
\usepackage{bbold}

\begin{document}

\newcommand{\be}{\begin{equation}}
\newcommand{\ee}{\end{equation}}
\newcommand{\bea}{\begin{eqnarray}}
\newcommand{\eea}{\end{eqnarray}}
\newcommand{\ra}{\rightarrow}
\newcommand{\callH}{\mathcal{H}}
\newcommand{\callF}{\mathcal{F}}
\newcommand{\ket}{\rangle}
\newcommand{\bra}{\langle}
\newcommand{\ha}{\hat{a}}
\newcommand{\hu}{\hat{u}}
\newcommand{\had}{\hat{a}^{\dagger}}
\newcommand{\hH}{\hat{H}}
\newcommand{\hV}{\hat{V}}
\newcommand{\hT}{\hat{T}}
\newcommand{\hW}{\hat{W}}
\newcommand{\hh}{\hat{h}}
\newcommand{\hn}{\hat{n}}
\newcommand{\hGamma}{\hat{\Gamma}}
\newcommand{\bh}{\mathbf{h}}
\newcommand{\bG}{\mathbf{G}}
\newcommand{\br}{\mathbf{r}}
\newcommand{\bz}{\mathbf{z}}
\newcommand{\bR}{\mathbf{R}}
\newcommand{\bx}{\mathbf{x}}
\newcommand{\by}{\mathbf{y}}
\newcommand{\bu}{\mathbf{u}}
\newcommand{\ba}{\mathbf{a}}
\newcommand{\bs}{\mathbf{s}}
\newcommand{\bk}{\mathbf{k}}
\newcommand{\bq}{\mathbf{q}}
\newcommand{\bp}{\mathbf{p}}
\newcommand{\bS}{\mathbf{\Sigma}}
\newcommand{\half}{\frac{1}{2}}
\newcommand{\prt}{\partial}
\newcommand{\ua}{\uparrow}
\newcommand{\da}{\downarrow}
\newcommand{\dr}{d^3}
\newcommand{\xc}{\textrm{xc}}
\newcommand{\x}{\textrm{x}}
\newcommand{\cc}{\textrm{c}}
\newcommand{\s}{\sigma}
\newcommand{\m}{{m}}
\newcommand{\drm}{d^{3m}}
\newcommand{\co}{\textrm{c}}
\newcommand{\Ha}{\textrm{H}}
\newcommand{\F}{\textrm{F}}
\newcommand{\Hxc}{\textrm{Hxc}}
\newcommand{\HK}{\textrm{HK}}
\newcommand{\LDA}{\textrm{LDA}}
\newcommand{\GGA}{\textrm{GGA}}
\newcommand{\GEA}{\textrm{GEA}}
\newcommand{\mom}{\hat{\mathbf{p}}}
\newcommand{\kf}{k_{\textrm{F}}}
\newcommand{\ef}{\epsilon_{\textrm{F}}}
\newcommand{\tpth}{(2 \pi)^3}
\newcommand{\bru}{\underline{\br}_m}
\newcommand{\bqu}{\underline{\bq}_m}
\newcommand{\dru}{d \bru}
\newcommand{\drd}{d^D}
\newcommand{\tpd}{(2 \pi)^D}
\newcommand{\tpt}{(2 \pi)^2}
\newcommand{\mf}{\mathcal{F}}
\newcommand{\llangle}{\langle \! \langle}
\newcommand{\rrangle}{\rangle \! \rangle}

\title{The density gradient expansion of correlation functions}

\author{Robert van Leeuwen}
\affiliation{Department of Physics, Nanoscience Center, University of Jyv\"askyl\"a, Survontie 9, 40014 Jyv\"askyl\"a, Finland}
\affiliation{European Theoretical Spectroscopy Facility (ETSF)}

\date{\today}

\begin{abstract}
We present a general scheme based on nonlinear response theory to calculate the expansion of correlation functions such as the pair-correlation function or
the exchange-correlation hole of an inhomogeneous many-particle system in terms of density derivatives of arbitrary order. 
 We further derive a consistency condition that is necessary for the
existence of the gradient expansion. This condition is used to carry out an infinite summation of terms
involving response functions up to infinite order from which it follows that
the coefficient functions of the gradient expansion can be expressed in terms the local density profile rather than 
the background density around which the expansion is carried out.
We apply the method to the calculation of the gradient expansion of the 
one-particle density matrix to second order in the density gradients and recover in an alternative manner 
the result of Gross and Dreizler (Z. Phys. A {\bf 302}, 103 (1981)) which was derived using the Kirzhnits method. 
The nonlinear response method is more general and avoids the turning point problem of the Kirzhnits expansion. We further give a description of
the exchange hole in momentum space and confirm the wave vector analysis of Langreth and Perdew (Phys. Rev. B{\bf 21}, 5469 (1980)) 
for this case. This is used to derive that the second order gradient expansion of the system averaged exchange hole
satisfies the hole sum rule and to calculate the gradient coefficient of the exchange energy without the need to regularize divergent integrals.

\end{abstract}
\maketitle

\section{Introduction}

Since the groundbreaking work of Hohenberg and Kohn \cite{HK} we know that the external potential of any
inhomogeneous quantum many-body system is a functional of its ground state density $n$. This implies that the
many-body ground state $|Ê\Psi [n] \rangle$ and hence any ground state expectation value
\be
A [n] = \langle \Psi [n] | \hat{A} | \Psi [n] \rangle
\label{Aop}
\ee 
for any operator $\hat{A}$ is a functional of the density. This idea has inspired an enormous amount
of work in a research field that is now known as density-functional theory. Density-functional theory 
\cite{DreizlerGross,Engelbook,Barthreview} is
currently one of the main approaches used in electronic structure theory. Over the years 
many extensions beyond its original formulation have been developed and currently it is widely applied in
solid state physics and quantum chemistry, both for ground state and time-dependent systems \cite{Ullrichbook}. 
Especially a large activity has gone into finding explicit expressions for the ground state energy $E [n]$ 
as a functional of the density for many-electron systems. The ground state energy is obtained by minimizing this functional
on an appropriate set of electronic densities, which is of great importance in determining structures and geometries
of molecules and solids.
By use of the Kohn-Sham method \cite{KS} this minimization problem is reduced to solving effective one-particle equations.
The construction of the effective potential in these equations
requires the knowledge of the so-called exchange-correlation energy functional $E_\xc [n]$. This functional
can be expressed as \cite{Almbladh,Gunnarsson}
\be
E_\xc [n] = \frac{1}{2}  \int d \br d \br' \frac{ n(\br)  \rho_{\xc} (\br' | \br)}{|\br -\br'|}
\label{lambint}
\ee
where $\rho_{\xc}(\br' |\br)$ is the coupling constant integrated exchange-correlation hole. The exchange-correlation hole has a physical interpretation as the difference
between the conditional and the unconditional probability (which is simply the density $n(\br')$) to find a electron at $\br'$, given that we know that
there is an electron at $\br$. Equation (\ref{lambint}) is an important relation
since it expresses the exchange-correlation energy in terms of a quantity that has a local physical interpretation
and can be studied by accurate wave function and many-body methods or modeled based on physical intuition.
It has therefore played an important role in the development of approximate density functionals \cite{Gunnarsson,Perdew:PRL85,PerdewWang:PRB86,Becke:JCP86,Becke:JCP88,BeckeRoussel,TschinkeZiegler}. 
One of the simplest and
already very successful approximations has been the local density approximation (LDA) in which the 
exchange-correlation hole is taken from the homogeneous electron gas and applied locally \cite{Gunnarsson} to systems with arbitrary density profiles. The accuracy of the
LDA has been considerably improved by means of the so-called generalized gradient approximations (GGAs) 
\cite{KurthPerdew}. A large class of commonly used GGAs is based on the so-called real-space cutoff
of the straightforward gradient expansion for the exchange and correlation hole. 
The first such functional for exchange
was proposed by Perdew \cite{Perdew:PRL85}. He noted that the gradient expansion of the exchange hole to second order
in the density gradients violates the negativity condition and the hole sum rule by which the exchange hole integrates to a deficit of one electron. 
By enforcing these physical constraints a density functional was obtained that greatly improved on the LDA for binding energies of molecules.
This procedure was later simplified \cite{PerdewWang:PRB86} using the fact that the exchange correlation energy (\ref{lambint}) can be written as
\be
E_\xc [n] = \frac{N}{2} \int d\by \frac{\langle \rho_{\xc} (\by) \rangle}{|\by|} \nonumber
\ee
where $N$ is the number of electrons in the system and 
\be
\langle \rho_{\xc} (\by) \rangle = \frac{1}{N} \int d\br \, n(\br) \rho_{\xc} (\br +\by | \br), 
\label{sa_hole}
\ee
is the so-called system-averaged exchange-correlation hole. This averaged hole still obeys the negativity
condition as well as the sum rule and can therefore be subjected to the real-space cut-off procedure. One advantage of
using the system averaged holes was to reduce the order of the derivatives in the gradient expansion. Furthermore it also
allowed for the real-space cut-off procedure to
be applied to the correlation hole since there is a known gradient expansion for the system averaged correlation hole
in the random phase approximation (RPA) but no known expansion for the correlation hole itself.
This was the basis for a GGA for the correlation energy \cite{Engelbook,Perdew:book,BPWbook,PBW:PRB96}. 
These GGAs relied heavily on the pioneering work by Langreth, Perdew and Mehl \cite{LangrethPerdew:PRB80,LangrethMehl:PRL81, LangrethMehl:PRB83}
who performed a wave-vector decomposition of the system averaged hole of Eq. (\ref{sa_hole}) and calculated the Fourier transform
\be
\langle \rho_{\xc} (\bk) \rangle = \int d\by \langle \rho_{\xc} (\by) \rangle e^{-i \bk \cdot \by} , \nonumber
\ee
from which the exchange correlation energy can be calculated as
\be
E_\xc [n] = \frac{N}{2} \int \frac{d \bk}{(2 \pi)^3} \langle \rho_{\xc} (\bk) \rangle \frac{4 \pi}{|\bk|^2} . \nonumber
\ee
We are not aware of any work that goes beyond references \cite{LangrethPerdew:PRB80,LangrethMehl:PRL81, LangrethMehl:PRB83} and 
improves on the straightforward gradient expansion of the system averaged correlation hole.
The knowledge on the gradient expansion of the correlation hole is very limited indeed. 
As mentioned before, in contrast to the work on the exchange hole
\cite{DreizlerGross,Perdew:PRL85}, there are no known expressions for the gradient expansion of the
non-system averaged correlation hole. \\
Part of the problem is that there has been no clear derivation on how to expand the expectation value of 
a general operator as in Eq.(\ref{Aop}) in terms of density gradients. A well-known expansion method for
a two-point function is the Kirzhnits expansion \cite{DreizlerGross} which is, however, specifically adapted to expanding the noninteracting one-particle
density matrix in powers of density gradients and can not be used for more general correlation functions. The first goal of this paper is, therefore, to present a scheme based on nonlinear response theory that can be used
to expand general correlation functions (such as the correlation hole) in terms of density gradients. \\
The second goal of the paper is to clarify a point that is often overlooked in carrying out gradient expansions.
We will illustrate the problem by considering the standard gradient expansion for a global quantity, namely the exchange-correlation energy $E_\xc$.
The starting point of any gradient expansion is the consideration of a density profile $n (\br)=n_0 + \delta n(\br)$ 
which consists of a small density variation $\delta n(\br)$ around a homogeneous density $n_0$. By considering the limit of
slow density variations one then finds that the lowest gradient correction to the exchange-correlation energy is an integral over a function of the form 
$B_\xc (n_0) (\nabla \delta n(\br))^2$ where the coefficient $B_\xc (n_0)$ is calculated from the static exchange-correlation
kernel of the homogeneous electron gas of density $n_0$ \cite{Barthreview}. Since $\nabla \delta n (\br) = \nabla n (\br)$
we can replace the density variation under the derivative operator by the full density profile $n(\br)$. However, the coefficient $B_\xc (n_0)$ still
depends on the background density $n_0$. This is a problem in application of the formula to general inhomogeneous systems such as molecules
or surfaces in which a background density cannot be unambiguously defined (assuming that a low order gradient expansion makes sense
in such systems \cite{SvendsenBarth:PRB96,Springer:PRB96}). The usual argument to get rid of the background dependence, is that the replacement
$B_\xc (n_0) \rightarrow B_\xc (n_0 + \delta n (\br))$ can be made since the difference between these two quantities is or order 
$\delta n(\br)$. However, it is not clear that this is consistent with gradient coefficients that arise from terms of order
$(\delta n(\br))^3$. The point was discussed clearly by Svendsen and von Barth \cite{SvendsenBarth:PRB96} who checked that 
the replacement of $n_0$ by the full density profile was consistent to third order in the density variation. This derivation 
was based on a specific relation between response functions that describe the change in the exchange energy to second and third order in the
 density variations. In reference \cite{AQC:2003} this derivation was extended by showing
that the replacement of $n_0$ by the full density profile is consistent to all orders in the density variation.
In this paper we will show that this is the case as well for the gradient expansion of correlation functions rather than scalar functions. This relies on
certain relations between the response functions that must be satisfied for the gradient expansion to exist.

The paper is divided as follows. In section \ref{gradexp} we derive the basic equations and consistency conditions for
the density gradient expansion of correlation functions. In section \ref{onemat} we derive the gradient expansion of the
one-particle density matrix for a noninteracting system with density $n(\br)$ (which is therefore equal to the density matrix of the
Kohn-Sham system) and discuss its symmetry properties. We further calculate the gradient expansion of the exchange-hole to
second order in the density gradients, both in real and in momentum space. The momentum space description is used to
demonstrate that the system averaged exchange hole satisfies the sum rule (but not the negativity condition), and to calculate the gradient
expansion of the exchange energy without the need to regularize divergent integrals. 
Finally in section \ref{conclusions} we present our conclusions and outlook.

\section{Gradient expansion of correlation functions}

\label{gradexp}

\subsection{Expansion in density variations}

Let us for a system with ground state density $n(\br)$ consider an arbitrary correlation function 
$F [n] (\br',\br)$.  Since the correlation function can be calculated from the many-body ground state, 
by virtue of the Hohenberg-Kohn theorem this function is a functional of the density. 
At this point it will not be important what the specific form of the correlation function is, as 
the structure of its gradient expansion will be independent of its specific form. 
Correlation functions of common interest are, for instance, the pair-correlation function, the exchange-correlation hole
or the one-particle density matrix.

When the density is constant in space, i.e. when $n(\br)=n_0$, we are describing the homogeneous
electron gas and due to translational and rotational invariance we have that $F [n](\br',\br) = F^0 (n_0, |\br-\br'|)$
where $F^0$ is a function of the homogeneous density and the distance between its spatial arguments.
When the density $n(\br) = n_0 + \delta n(\br)$ deviates from the constant density $n_0$ by a small amount $\delta n(\br)$
we can expand $F$ in powers of this density variation. To derive compact expressions
we first introduce the notation
\bea
\bru &=& (\br_1, \ldots, \br_m) \nonumber \\
\dru &=& d\br_1 \ldots d \br_m \nonumber
\eea
for the collection of $m$ position vectors and the corresponding integration volume
elements. We further define
\bea
\delta n (\bru ) &=& \prod_{j=1}^m \delta n (\br_j) \nonumber .
\eea
With these conventions the expansion of $F$ in density variations is given by
\bea
\lefteqn{ F[n] (\br',\br) = F^0(n_0, |\br'-\br|) + }\nonumber \\
&&\sum_{m=1}^{\infty} \frac{1}{m!} \int \dru \,  F^\m (n_0;\br', \br, \bru) \delta n (\bru) \nonumber
\eea
where we defined
\be
F^\m (n_0;\br' , \br, \bru ) = \frac{\delta^m F(\br',\br)}{\delta n(\br_1) \ldots \delta n(\br_m)}|_{n_0} .
\label{Fdef}
\ee
The $n_0$ dependence of the functions $F^m$ will be important
for the gradient expansion, but to shorten notation we suppress this dependence in the notation for the time being
and will reintroduce it later.
Here we assume that the derivatives $F^m$ exist, which means that we assume that $F$ is a smooth functional
of the density. In the electron gas two Taylor expansions around two different
values of $n_0$ have very likely the same mathematical convergence properties
since the two systems have the same physical properties (unless we are close to a very low density 
corresponding to a Wigner crystal transition). The point $n_0$ is therefore unlikely to be a 
non-analytic point. When the derivatives $F^m$ do exist we can therefore expect the Taylor series (\ref{Fdef})
to converge whenever $|\delta n(\br)| /n_0 \ll 1$ or when they are integrable with a small integral norm. 
The density variations are therefore assumed to be small
but we do not need to put any constraint on how rapid they can vary in space and 
therefore their spatial derivatives can be large. 
We note, however, that the notion of the existence of the functional derivative is closely related to
the allowed domain of densities and the norm defined on this function space. 
The smallness or integrability of $|\delta n(\br)|$ implies the use of a supremum norm or possibly an
$L^p$-integral norm (i.e. requiring $|\delta n(\br)|^p$ to be integrable).
Other norms may induce weaker constraints on the derivative functions $F^m$ but stronger constraints on the density
variations $\delta n(\br)$. A rigorous discussion on the issue of functional differentiability for the case of the
energy functional is given in Refs.\cite{Lammert1,Lammert2,Eschrig1,Eschrig2}.

Let us now look at the
symmetry properties of the functions $F^m$. 
As follows directly from the definition (\ref{Fdef}) of the functions $F^\m$ and the assumption of differentiability,
we have the permutational symmetry
\be
F^\m (\br',\br, \br_1 \ldots \br_m) = F^\m (\br',\br, \br_{\pi(1)} \ldots \br_{\pi(m)}) \nonumber
\ee
for all permutations $\pi$ of the integers $1 \ldots m$.
Since the functions $F^\m$ are evaluated at the homogeneous density $n_0$ they also have the
spatial symmetry properties of the homogeneous electron gas. 
These symmetries are the translational symmetry over an arbitrary vector $\mathbf{a}$, rotational
symmetry under arbitrary rotations by a rotation matrix $R$, and inversion symmetry. 
If we denote by
\bea
\underline{\mathbf{a}} &=& (\mathbf{a}, \ldots , \mathbf{a}) \nonumber \\
R\bru  &=& (R \br_1, \ldots, R \br_m) \nonumber
\eea
the $m$-tuples of translation vectors and rotated position vectors we can write
\bea
F^\m ( \br',\br, \bru) &=& F^\m (\br'+\mathbf{a},\br+\mathbf{a}, \bru + \underline{\mathbf{a}}) 
\label{eq:mtrans}\\
F^\m ( \br',\br,\bru) &=& F^\m (R\br',R\br, R\bru ) 
\nonumber \\
F^\m ( \br',\br,\bru) &=& F^\m (-\br',-\br, -\bru) .
\nonumber
\eea
Since in Eq.(\ref{eq:mtrans}) the vector $\ba$ is arbitrary we can in particular choose $\ba=-\br$
and define a new function $N^m$ depending on $m+1$ independent vectors by the relation
\begin{align}
F^\m (\br',\br, & \,\br_1 \ldots \br_m)   \nonumber \\
& = F^\m (\br'-\br,0, \br_1-\br, \ldots ,\br_m-\br) \nonumber \\
& = N^\m ( \br'-\br,\br_1-\br, \ldots ,\br_m-\br) .
\label{Ndef}
\end{align}
The difference vector $\br'-\br$ will appear several times in our equations and
it will therefore be convenient to define the short notation $\by = \br' -\br$.
We will further use $y=|\by|$ to denote the length of this vector. 
Our interest will be in the functions $N^\m$ and in particular their Fourier transforms
\be
N^\m (\by,\bqu) =
\int \dru \, N^\m ( \by,\bru) \, e^{-i \bq_1 \cdot \br_1 - \ldots - i \bq_m \cdot \br_m}
\label{Nfourier}
\ee
with Fourier inverse
\be
N^\m (n_0; \by,\bru) =
\int \frac{d \bqu}{(2 \pi)^{3m}} \, N^\m (\by,\bqu)  e^{i \bq_1 \cdot \br_1 + \ldots + i \bq_m \cdot \br_m} . \nonumber
\ee
With these definitions we obtain the following expansion for $F$ in powers of the density variations
\begin{widetext}
\bea
F[n] (\br',\br) &=& F^0(n_0, y) + 
\sum_{m=1}^{\infty} \frac{1}{m!} \int \dru \, 
\int \frac{d \bqu}{(2 \pi)^{3m}} \, N^\m (\by,\bqu)  e^{i \bq_1 \cdot (\br_1-\br) + \ldots + 
i \bq_m \cdot (\br_m-\br) } \delta n (\bru) \nonumber \\
&=& F^0(n_0, y) + 
 \sum_{m=1}^{\infty} \frac{1}{m!} 
\int \frac{d\bqu}{(2 \pi)^{3m}} \, N^\m (\by,\bqu)  e^{-i (\bq_1 + \ldots +\bq_m) \cdot \br } 
\delta n (-\bqu)
\label{eq:fexp}
\eea
\end{widetext}
This equation forms the basis of the gradient expansion. To proceed we further
derive a consistency condition that is necessary for the existence of the gradient expansion.
It follows directly from the definition (\ref{Fdef}) of the functions $F^\m$ that
\be
\delta F^\m (\br',\br,\bru) = \int d\br'' \, F^{m+1}(\br',\br,\br'',\bru) \delta n(\br'') . \nonumber
\ee
Taking $\delta n(\br'')=\delta n_0$ then to be a uniform shift in the density (which means that we
look at a compression or decompression of the electron gas) yields
\be
\frac{\prt F^\m }{\prt n_0} (\br',\br,\bru)= \int d\br'' \, F^{m+1}(\br',\br,\br'',\bru) . \nonumber
\ee
We can translate this condition to a condition on $N^\m$ using its definition (\ref{Ndef}) and the definition (\ref{Nfourier}) of
its Fourier transform. This gives the condition
\be
\frac{\prt N^\m }{\prt n_0} (\by,\bq_1 \ldots \bq_m)=  N^{m+1}(\by,0,\bq_1 \ldots \bq_m) .
\label{eq:Ncond}
\ee
This relation is of key importance for the existence of the gradient expansion. Without it we would not be
able to eliminate the dependence of the gradient coefficients on the homogeneous density $n_0$ in favor of a dependence
on the actual density profile. As we will see, this requires a summation over response functions $F^m$ to infinite order.
There is another important advantage of this re-summation. It will allow us to relax the constraint that the density
variations be small (but varying arbitrarily fast) and replace it with the constraint that the density variations vary
slowly but with an arbitrary amplitude. This is discussed in the next section.

\subsection{The gradient expansion}

In order to expand $F$ in density gradients we have to assume that the density gradients are small. This 
constraint is most easily phrased in momentum space by requiring that the Fourier coefficients $\delta n(\bq)$
of the density variation have their main contribution from small wave vectors $\bq$. If this is the case then the main 
contribution to the integrals in Eq.(\ref{eq:fexp}) comes from this region of small vectors and we
can expand the functions $N^m$ in powers of the wave vectors. 
Subsequently carrying out the integrals in Eq.(\ref{eq:fexp})
then leads to an expansion in density gradients, since powers in wave vectors correspond to orders of derivatives in real space.
Since the response functions $N^m$ typically vary very
rapidly for wave vectors close to the Fermi surface (or multiples thereof) we require that the Fourier coefficients $\delta n(\bq)$ of the  
density variation have their main contributions from wave vectors that satisfy $|\bq| < \kf$ where $\kf$ is the
Fermi wave vector. Note that the procedure of interchanging integration and summation requires absolute convergence
of the wave vector expansion. It is hard to prove this property for the general response functions that we consider here
and we therefore have to assume its validity.

The next task is then to expand the functions $N^m$ in powers of wave vectors. This task is simplified 
when we exploit the symmetry of the these functions. As follows directly from Eqs.(\ref{Ndef}) and (\ref{Nfourier})
the functions $N^\m$ in Fourier space inherit the symmetry properties of $F^\m$. They are symmetric functions
of their wave vectors, i.e.
\be
N^\m ( \by,\bq_1 \ldots \bq_m) = N^\m ( \by, \bq_{\pi(1)}  \ldots \bq_{\pi(m)} ) \nonumber
\ee
for any permutation $\pi$ of the integer labels, and are invariant under rotations and
inversion 
\bea
N^\m ( \by,\bq_1 \ldots \bq_m) &=& N^\m ( R\by,R \bq_1  \ldots R \bq_m ) 
\nonumber \\
N^\m ( \by,\bq_1 \ldots \bq_m) &=& N^\m ( -\by,- \bq_1  \ldots - \bq_m ) .
\nonumber 
\eea
From these equations we therefore see that any power expansion of the functions $N^m$ must lead to an
expansion in terms of the spatial vector $\by$ and the wave vectors $\bq_j$ which is invariant under rotations and inversions of these vectors.
The mathematical question is then which polynomials have these properties. This was answered in the classic book by Weyl \cite{Weyl:book}; 
every such polynomial must be a function of the variables
$\by \cdot \by$, $\by \cdot \bq_i$ and $\bq_i \cdot \bq_j$. We see that these inner products are indeed invariant under rotations and
inversions. Since any polynomial in powers of $y^2=\by \cdot \by$ can be re-summed to a function of $y$ we find that the general
expression for $N^m$ to second order in the wave vectors $\bq_j$ is given by
\begin{widetext}
\bea
N^\m (\by,\bq_1 \ldots \bq_m) &=& 
 N_0^\m (y) + N_1^\m (y) \, \by \cdot  \sum_{i=1}^m \bq_i 
  +  N_2^\m (y) \sum_{i=1}^m \bq_i^2   + N_3^\m (y) \, \sum_{i=1}^m (\by \cdot \bq_i)^2 \nonumber \\
 && + N_4^\m (y) \, \sum_{i>j}^m (\bq_i \cdot \bq_j) + 
 N_5^\m (y) \, \sum_{i>j}^m (\by \cdot \bq_i) (\by \cdot \bq_j)+ \ldots
 \label{eq:Nexp}
\eea
\end{widetext}
where $\bq_i^2=\bq_i \cdot \bq_i$ and where we took into account that this expansion must be invariant under permutations of the wave vectors $\bq_j$.
This expansion is readily extended to higher order powers in the wave vectors. 
For a given choice of correlation function $F$ the practical task will be to determine the explicit form
of the coefficient functions $N_j^m (y)$ as a function of $y$ and the background density $n_0$.
If we insert Eq.(\ref{eq:Nexp}) into Eq.(\ref{eq:fexp}) and Fourier transform back to real space we obtain the expansion
\bea
 F (\br',\br) = F^0(n_0, y) + \sum_{m=1}^\infty \frac{1}{m!}  \sum_{j=0}^\infty N_j^\m (y) A_j^\m (\by,\br) .
 \label{eq:fexp2}
\eea
Let us analyze the explicit form of the first six coefficients $A_j^m$ for $j=0,..,5$ of this expansion, since
these are exactly the terms that we need for a gradient expansion to second order in the density derivatives.
The first term for $j=0$ is simply given by
\be
A_0^\m = \delta n(\br)^m . \nonumber
\ee
The terms with $j=1,2,3$ in Eq.(\ref{eq:fexp2}) involve according to Eq.(\ref{eq:Nexp}) the Fourier transforms of $m$ symmetry equivalent terms and
acquire the following forms in real space
\bea
A_1^\m &=&  i \, m ( \delta n(\br))^{m-1} \, \by \cdot \nabla  \delta n(\br) \nonumber \\
A_2^\m &=& -m  ( \delta n(\br))^{m-1} \, \nabla^2  \delta n(\br) \nonumber \\
A_3^\m &=& -m  ( \delta n(\br))^{m-1} y^2 (\frac{\by}{y} \cdot \nabla)^2 \delta n(\br) \nonumber 
\eea
In the derivation of the last term we
 used that for an arbitrary function $f$
\be
(\frac{\by}{y} \cdot \nabla)^2 f = \sum_{ij} \frac{y_i}{y}  \prt_i (\frac{y_j}{y} \prt_j) f(\br) = \sum_{ij} \frac{y_i y_j}{y^2} \prt_i \prt_j f
\nonumber
\ee
where with $\br=(x_1,x_2,x_3)$ we used the notation $\partial_i=\partial/\partial x_i$ as well as $y_i=x_i'-x_i$. This
expression is readily checked using
\be
\prt_i ( \frac{y_j}{y}  ) = -\frac{\delta_{ij}}{y} + \frac{y_i y_j}{y^3} . \nonumber
\ee
Finally the terms with $j=4,5$ in Eq.(\ref{eq:fexp2}) involve according to Eq.(\ref{eq:Nexp}) the Fourier transforms
of $m(m-1)/2$ symmetry equivalent terms which yields
\bea
A_4^\m &=& - \half m(m-1) ( \delta n(\br))^{m-2} \, (\nabla  \delta n(\br))^2 \nonumber \\
A_5^\m &=& - \half m(m-1) ( \delta n(\br))^{m-2} \, (\by \cdot \nabla  \delta n(\br))^2 . \nonumber
\eea
We see that a general coefficient function $A_j^m$ depends on the density variation and gradients thereof.
The functions $\delta n (\br)$ that appear under the gradient operators can be replaced by the actual
density profile $n(\br) = n_0 + \delta n(\br)$. However, we see that we are still left with powers
of $\delta n (\br)$ which are unprotected by derivative operators and which therefore can not be replaced by the full density profile $n(\br)$. It is 
precisely at this point that the consistency conditions
(\ref{eq:Ncond}) play a crucial role. 
If we use these conditions in Eq.(\ref{eq:Nexp}) then we see that
\be
N_j^{m+1}(y)=\frac{\prt N_j^\m}{\prt n_0} (y) .
\label{eq:Ncond2}
\ee
This equation relates certain coefficients coming from higher order response functions $N^m$
to those of lower order ones.
In particular it tells us that by iteration
\bea
N_0^\m (y) &=& \frac{\prt^\m F^0 }{\prt n_0^m} (y) \nonumber \\
N_j^\m (y) &=& \frac{\prt^{m-1} N_j^{1} }{\prt n_0^{m-1}} (y) \quad (j=1,2,3)  \nonumber \\
N_j^\m (y) &=& \frac{\prt^{m-2}  N_j^{2}}{\prt n_0^{m-2}} (y)  \quad (j=4,5) . \nonumber
\eea
If we insert these expressions together with the explicit form of the coefficient functions $A_j^m$
back into Eq.(\ref{eq:fexp2})
in Eq.(\ref{eq:fexp2}) and shift some indices we obtain the expansion
\begin{widetext}
\bea
 F (\br',\br) &=& F^0 ( y) +
\sum_{m=1}^{\infty} \frac{1}{m!} \frac{\prt^m F^0 }{\prt n_0^m}   (y) \delta n(\br)^m + \nonumber \\
 &&  \sum_{m=0}^{\infty} \frac{1}{m!} \delta n(\br)^m \big[   i  \frac{\prt^{m} N_1^{1} }{\prt n_0^{m}}
 (\by \cdot \nabla  n(\br))  -    \frac{\prt^{m} N_2^{1} }{\prt n_0^{m}}  \nabla^2 n(\br)
 -   \frac{\prt^{m} N_3^{1} }{\prt n_0^{m}} y^2 \, (\frac{\by}{y} \cdot \nabla)^2 n(\br)\big]  +\nonumber \\
 &&  - \frac{1}{2}\sum_{m=0}^{\infty} \frac{1}{m!} \delta n(\br)^m \big[ \frac{\prt^{m} N_4^{2} }{\prt n_0^{m}}  
 (\nabla n(\br))^2 + \frac{\prt^{m} N_5^{2} }{\prt n_0^{m}}  (\by \cdot \nabla n(\br))^2 \big] + \ldots
\eea
In this expression we recognize several Taylor series of the coefficient functions $N_j^m (n_0+\delta n(\br), y)$ in powers of $\delta n(\br)$.
These Taylor series can be re-summed to finally give
\bea
F (\br',\br) &=& F^0 (n(\br), y) + 
 i N_1^{1} (n(\br),y) \, \by \cdot \nabla n(\br) - N_2^{1} (n(\br),y) \, \nabla^2 n(\br)
- N_3^{1} (n(\br),y) \, y^2 ( \frac{\by}{y} \cdot \nabla)^2 n(\br) \nonumber \\
&&  - \half N_4^{2} (n(\br),y) \, (\nabla n(\br))^2  - \half N_5^{2} (n(\br),y) \, (\by \cdot \nabla n(\br))^2 + \ldots ,
\label{f_exp}
\eea
where the implicit dependence on $n_0$ of the coefficient functions is now replaced by a dependence on the full density profile. We see that this is achieved by
summing over response functions $N^m$ to infinite order.
For clarity we reinstated the explicit density dependence of the coefficients in Eq.(\ref{f_exp}) to indicate that they are local functions
of the density $n(\br)$ and therefore have a nontrivial spatial dependence on both $\br$ and $y$.
We stress again that the derivative condition (\ref{eq:Ncond}) was essential in eliminating the dependence of the gradient coefficient
on the reference density $n_0$ in favor of a dependence on the full density profile. Having obtained the general form of gradient expansion
we can wonder about its convergence. In order for the gradient expansion to be useful it at least needs to be an asymptotic series. 
This is known to be the case for the gradient expansion of some commonly studied functionals. For instance, for small densities variations
very accurate results for the exchange energy are obtained by summing all terms up to fourth order in the density derivatives \cite{Barthreview,Springer:PRB96}.

Having obtained the general gradient expansion Eq.(\ref{f_exp})
it remains to find explicit expressions for the coefficient functions $N_j^m$. We will describe how to do this in detail for the coefficients
needed for a gradient expansion to second order in the density derivatives.
To calculate the coefficients $N_1^{1},N_2^{1}$ and $N_3^{1}$ we need to calculate the
function $N^{1}(n_0,\by,\bq)$ for the homogeneous electron gas and expand it in powers of $\bq$:
\be
N^{1}(n_0,\by,\bq) = N_0^{1} (n_0,y) + N_1^{1} (n_0,y)  \by \cdot \bq + N_2^{1} (n_0,y) \bq^2 +
N_3^{1} (n_0,y) (\by \cdot \bq)^2 + \ldots  
\label{N1_expansion}
\ee
The determination of the coefficients $N_4^{2}$ and  $N_5^{2}$ requires knowledge of the second order
response 
function
\bea
N^{2}(n_0,\by,\bq_1,\bq_2) &=&N_0^{2} (n_0, y) + N_1^{2} (n_0 ,y)  \by \cdot (\bq_1 + \bq_2) 
+
 N_2^{2} (n_0, y) (\bq_1^2 + \bq_2^2) + \nonumber \\
&& N_3^{2} (n_0, y) ((\by \cdot \bq_1)^2 + (\by \cdot \bq_2)^2 ) 
+ N_4^{2} (n_0, y)  (\bq_1 \cdot \bq_2) + N_5^{2} (n_0,y) (\by \cdot \bq_1) (\by \cdot \bq_2) + \ldots 
\label{N2_expansion}
\eea
\end{widetext}
The expression (\ref{f_exp}) together with Eqs.(\ref{N1_expansion}) and (\ref{N2_expansion}) determine completely
the gradient expansion of an arbitrary correlation function $F$ to second order in the density gradients.
These equations form the main result of the present work. If expansions to higher order gradients are
required they can be derived without difficulty along the lines described above. To do this one needs to construct higher order symmetric polynomials
in the wave vectors and carry out the required Fourier transforms. The consistency conditions Eq.(\ref{eq:Ncond}) or equivalently Eq.(\ref{eq:Ncond2}) then
allow for a complete re-summation and leads to gradient coefficients depending on $n(\br)$.
What is needed to determine the explicit form of the gradient coefficients in practice is the determination
of the functions $N^m$. How to do this for $N^1$ and $N^2$ is described in the next section.
 
\subsection{Determination of $N^1$ and $N^2$}

A practical calculation of the coefficients of the gradient expansion of Eq.(\ref{f_exp}) requires
the knowledge of the the functions $N^{1}$ and $N^{2}$. According to Eqs.(\ref{Fdef}) and (\ref{Ndef}) these are defined as
the first and second density derivatives of the correlation function $F$ with respect to the density, i.e.
\bea
\frac{\delta F (\br',\br)}{\delta n (\br'')} |_{n_0} &=&  N^{1}(\by,\br''-\br)  \nonumber \\
\frac{\delta F (\br',\br)}{\delta n (\br'') \delta n(\br''')} |_{n_0} &=&  N^{2}(\by,\br''-\br,\br'''-\br) . \nonumber
\eea
To derive useful expressions for these functions it is convenient to transform derivatives with respect to the density to derivatives with
respect to the external potential as such derivatives appear naturally in perturbation theory. We can then use the well-developed
tools of perturbation theory to calculate these functions. Let us therefore
define the new functions
\bea
\mathcal{F}^1 (\by,\br''-\br) &=& \frac{\delta F (\br',\br)}{\delta v (\br'')}|_{n_0} 
\label{mathf1}\\
\mathcal{F}^2 (\by,\br''-\br,\br'''-\br) &=& \frac{\delta^2 F (\br',\br)}{\delta v (\br'') \delta v (\br''')} |_{n_0} .
\label{mathf2}
\eea
Since we evaluate these functions for the homogeneous electron gas they only depend on
differences between the spatial coordinates. It will therefore be convenient to also define their Fourier transforms by
\bea
\mathcal{F}^1 (\br_1,\br_2 ) &=& \int \frac{d \bp d \bq}{(2 \pi)^6} \mf^1 (\bp,\bq) e^{i \bp \cdot \br_1 + i \bq \cdot \br_2} \nonumber \\
\mathcal{F}^2 (\br_1,\br_2,\br_3) &=& \int \frac{d \bk d \bp d \bq }{(2 \pi)^9} \mf^2 (\bk,\bp,\bq)  \nonumber\\
&& \times e^{i \bk \cdot \br_1 + i \bp \cdot \br_2  + i \bq \cdot \br_3} \nonumber 
\eea
as well as their partial Fourier transforms
\begin{align}
\mf^1 (\by,\bq) &= \int \frac{d \bp}{(2 \pi)^3} \mf^1 (\bp,\bq) e^{i \bp \cdot \by } \nonumber \\
\mf^2 (\by,\bp,\bq) &= \int \frac{d \bk}{(2 \pi)^3} \mf^2 (\bk,\bp,\bq) e^{i \bk \cdot \by } . \nonumber
\end{align}
The chain rule of differentiation gives a connection between the derivatives with respect to the density and the
derivatives with respect to the potential of the correlation function $F$. We have
\begin{align}
 \frac{\delta F (\br_1,\br_2)}{\delta v (\br_3)} &= \int d \br_4 \, 
\frac{\delta F (\br_1,\br_2)}{\delta n(\br_4)} \frac{\delta n(\br_4)}{\delta v (\br_3)} \label{rule1} \\
 \frac{\delta^2 F (\br_1,\br_2)}{\delta v (\br_3) \delta v (\br_4)} &= 
\int d \br_5 \, 
\frac{\delta F (\br_1,\br_2)}{\delta n(\br_5)} \frac{\delta^2 n(\br_5)}{\delta v (\br_3) \delta v (\br_4)}
 \nonumber \\
+ \int d \br_5 d \br_6 & \,
\frac{\delta^2 F (\br_1,\br_2)}{\delta n(\br_5) \delta n(\br_6)} \frac{\delta n(\br_5)}{\delta v (\br_3)}
\frac{\delta n(\br_6)}{\delta v (\br_4)} .\label{rule2}
\end{align}
We see that in these expressions the first and second order derivative of the density
with respect to the potential appear. 
We therefore define
the linear density response function by
$\chi$ 
\begin{align}
\frac{\delta n (\br_1)}{\delta v(\br_2)} |_{n_0}   =   \chi (\br_2-\br_1)
= \int \frac{d \bq}{(2 \pi)^3} \chi (\bq) e^{i \bq \cdot (\br_2-\br_1)}  \nonumber
\end{align}
as well as the second order density response function $\chi^2$ by
\begin{align}
 \frac{\delta^2 n (\br_1)}{\delta v(\br_2) \delta v (\br_3) } |_{n_0} &= \chi^2 (\br_2-\br_1,\br_3-\br_1)  \nonumber \\ 
  =  & \int \frac{d \bp d \bq}{(2 \pi)^6}  \chi^2 (\bp,\bq) e^{i \bp \cdot (\br_2-\br_1) + i  \bq \cdot (\br_3-\br_1)} . \nonumber
\end{align}
Using these definitions we find
by Fourier transforming Eqs.(\ref{rule1}) and (\ref{rule2}) that
\begin{align}
N^1 (\by,\bq) & =  \frac{\mf^1 (\by,\bq)}{ \chi (\bq)}
\label{n1exp} \\
N^2 (\by,\bp,\bq)  & = 
  \frac{ \big[ \mf^2 (\by,\bp,\bq) - 
 N^1 (\by,\bp+\bq) \chi^2 (\bp,\bq) \big] 
 }{\chi (\bp) \chi (\bq)}
 \label{n2exp}
\end{align}
These equations give the desired relation between the density derivatives $N^1$ and $N^2$ and the potential derivatives
$\mf^1$ and $\mf^2$ of the correlation function $F$. Relations between the higher order derivatives $N^m$ and $\mf^m$ can be
derived in a completely analogous way.
From Eqs.(\ref{n1exp}) and (\ref{n2exp}) we see that to calculate the functions $N^1$ and $N^2$, and hence the
gradient expansion coefficients of $F$ to second order in the gradients, we need to calculate the
density response and the change in $F$ to second order in a perturbing potential $\delta v$.
The problem is thereby reduced to doing perturbation theory on the electron gas. This is a well-developed
field of research. One option is to use diagrammatic perturbation theory \cite{ourbook}. In the following section
we will use standard perturbation theory to obtain these response function for the case that $F$
is the one-particle density matrix and use that to calculate the gradient expansion of the exchange hole.
\\\\

\section{Gradient expansion of the one-particle density matrix and the exchange hole}

\label{onemat}

\subsection{The one-particle density matrix}

As an application of our formalism we carry out the gradient expansion of the one-particle
density matrix for a noninteracting 
system with density $n(\br)$. This problem has received large interest in the past since 
both the exchange energy $E_{\rm x} [n]$ as well the Kohn-Sham kinetic energy $T_s [n]$ 
are explicit functionals of such a noninteracting density matrix \cite{DreizlerGross,Engelbook}.
Therefore a gradient expansion of this density matrix directly leads to a gradient expansion of
these two functionals. Such a gradient expansion was first carried out by Gross and Dreizler \cite{GrossDreizler:81}
using the Kirzhnits expansion \cite{DreizlerGross,Esa}. This expansion is adjusted to the specific form of the noninteracting
one-particle density matrix and can not be generalized to arbitrary correlation functions.
The method presented in this work can, on the other hand, be applied to arbitrary correlation functions, but to demonstrate
the soundness of our formalism we will use it to provide an alternative derivation of the result
obtained earlier by Gross and Dreizler using the Kirzhnits expansion. An advantage of our derivation based on nonlinear
response theory is that
it avoids the turning point problem encountered in the Kirzhnits approach \cite{DreizlerGross}.

Let us start by defining the one-particle density matrix in second quantization as
\be
\gamma (\br',\br) = \sum_{\sigma} \langle \Psi | \hat{\psi}^\dagger (\br \sigma) \hat{\psi} (\br' \sigma)Ê| \Psi  \rangle \nonumber
\ee
where $\sigma$ is a spin coordinate. We will consider the case of spin-compensated systems.
Then, in a noninteracting system with $2 N$ electrons, the density matrix is given in terms of
one-particle wave functions $\varphi_j (\br)$ by
\be
\gamma (\br', \br) = \sum_{j=1}^N 2 \, \varphi_j (\br') \varphi_j^* (\br)
\label{onemat2}
\ee
where the pre-factor $2$ is a spin factor.
The one-particle states are eigenstates of a one-particle
Schr\"odinger equation
\be
\left( - \frac{1}{2} \nabla^2 + v (\br) \right) \varphi_j (\br) = \epsilon_j \varphi_j (\br)
\label{schr}
\ee
where $v(\br)$ is the external potential. 
To determine the gradient expansion coefficients
of $\gamma$ we need to determine how the density matrix changes if we
make small changes $v(\br) \rightarrow v(\br)+ \delta v(\br)$ in the
external potential. More precisely, we need to calculate
the functional derivatives 
\begin{align}
\gamma^1 (\br_1, \br_2,\br_3) &=\frac{\delta \gamma (\br_1,\br_2)}{ \delta v (\br_3)} \nonumber \\
\gamma^2 (\br_1, \br_2,\br_3, \br_4) &= \frac{\delta^2 \gamma (\br_1,\br_2)}{\delta v(\br_3)  \delta v (\br_4)} \nonumber
\end{align}
which are the equivalents of the functions $\mf^1$ and $\mf^2$ of Eqs.(\ref{mathf1}) and
(\ref{mathf2}) for the specific choice of correlation function $F=\gamma$ (note that they become functions of two and three independent vectors
respectively when evaluated for a homogeneous system). To do this it is sufficient to know the
functional derivatives of the orbitals $\varphi_j$ and eigenvalues $\epsilon_j$ with respect to the potential $v$. These
are simply given by doing first order perturbation theory on the one-particle Schr\"odinger Eq.(\ref{schr}). We find
\begin{align}
\frac{\delta \epsilon_j}{\delta v(\br) } & = |\varphi_j (\br)|^2 
\label{epsderv}\\
\frac{\delta \varphi_j (\br)}{\delta v(\br') } & = \sum_{k \neq j}^\infty 
\frac{\varphi_k (\br) \varphi_k^* (\br') \varphi_j (\br')}{\epsilon_j - \epsilon_k} .
\label{phiderv}
\end{align}
With these relations we can differentiate Eq.(\ref{onemat2}) twice with respect to
the potential. Afterwards we then need to insert the plane wave states $\varphi_{\bk} (\br)$
and their eigenvalues $\epsilon_{\bk}$ appropriate for the homogeneous electron gas into the final expressions.
In order not to interrupt the presentation these calculations are given in Appendix \ref{gammaderv} and we simply present the
results of these calculations here. 
If we define the Fermi factors by
\be
n_\bk = \theta (\kf - |\bk|) , \nonumber
\ee
where $\theta$ is the Heaviside function and where
$\kf = (3 \pi^2 n_0)^{1/3}$ is the Fermi wave vector, then
the resulting expressions are given by
\begin{align}
\gamma^{1} (\by, \bq) & = 2 \int \frac{d \bp}{(2 \pi)^3} \frac{n_{\bp+\bq}-n_\bp}{\epsilon_{\bp+\bq}-\epsilon_\bp} 
\, e^{i \bp \cdot \by}
\label{gamma_1} \\
\gamma^{2} (\by,\bp,\bq)
& = 2 \int \frac{d \bk}{(2 \pi)^3}\, e^{i \bk \cdot \by} [ \Phi (\bk,\bk +\bp + \bq,\bk + \bp)  \nonumber \\
&  + \Phi (\bk,\bk +\bp + \bq,\bk + \bq) ] \label{gamma_2}
\end{align}
where
\be
\Phi (\bk,\bp,\bq) =  \frac{1 }{\epsilon_\bq-\epsilon_\bp}  \left(     \frac{ n_\bq-n_\bk} {\epsilon_\bq - \epsilon_\bk } -  
\frac{ n_\bp- n_\bk} {\epsilon_\bp - \epsilon_\bk }        \right) .
\ee
In this expression $\epsilon_{\bk} = |\bk|^2 /2$ are the one-particle energies. 

It remains to calculate the first and second order
density response functions $\chi (\bq)$ and 
$\chi^2 (\bp,\bq)$ to evaluate the functions $N^1$ and $N^2$ of Eqs. (\ref{n1exp}) and (\ref{n2exp}). 
Fortunately, for the specific choice $F=\gamma$ that we made these density response functions are simply
given by 
\begin{align}
\chi (\bq) &= \gamma^1 (\by=0,\bq) \label{gamexp1}\\
\chi^2 (\bp,\bq) &= \gamma^1 (\by=0,\bp,\bq) \label{gamexp2}.
\end{align}
They are therefore obtained directly from Eqs. (\ref{gamma_1}) and (\ref{gamma_2})
so that we do not need to do any extra work to calculate them.

To calculate the gradient coefficient to second order
in the gradients we now need to expand $\gamma^1$ and $\gamma^2$ to second order in the wave vectors
$\bp$ and $\bq$. Since these calculations are somewhat lengthy we do not present them here and refer to Appendix \ref{N1N2} 
for a description of the main steps. The resulting expansions are given by
\begin{widetext}
\bea
\gamma^{1} (\by,\bq) &=& - \frac{\kf}{\pi^2} \left(   \frac{\sin z}{z}  \left[ 1 - \frac{i \bq \cdot \by}{2} -  \frac{(\bq \cdot \by)^2}{6}\right]
   - \frac{\bq^2}{12 \kf^2}   \cos (z) \right)  \label{gam1_result}\\
\gamma^{2} (\by,\bp,\bq) &=& \frac{1}{\pi^2 \kf} 
\left( \cos (z) - \frac{i}{2} (\bp + \bq) \cdot \by \cos(z) 
+ \frac{1}{12 \kf^2} (\bp^2+\bq^2 + \bp \cdot \bq) (\cos (z) + z \sin(z)) \right. \nonumber \\
&& \left. - \frac{1}{24} [  (\bp \cdot \by)^2 + (\bq \cdot \by)^2 + 3 ((\bp+ \bq) \cdot \by)^2 Ê] \cos (z) \right)
\label{gam2_result}
\eea
where we defined $z=\kf y$.
The expansions for the first and second order response functions $\chi$ and $\chi^2$ of Eqs.(\ref{gamexp1}) and (\ref{gamexp2}) are 
obtained by taking $\by=0$ in these expressions. Together which Eqs.(\ref{gam1_result}) and (\ref{gam2_result}) these expression 
then immediately determine the functions $N^{1}$ and $N^{2}$ from Eqs.(\ref{n1exp}) and (\ref{n2exp}). 
The final result of this calculation to second order in the wave vectors is given by
\bea
N^{1} (\by, \bq) &=&
 \frac{\sin z}{z} \left[ 1 - \frac{i \bq \cdot \by}{2 } 
- \frac{(\bq \cdot \by)^2}{6} \right]+ \frac{\bq^2}{12 \kf^2} \frac{ \sin (z) - z \cos (z) }{z} \nonumber \\
N^{2} (\by,\bp,\bq) &=& \frac{\pi^2}{\kf^3} 
\left[ \left( \cos (z) - \frac{\sin (z)}{z} \right) \left( 1 - \frac{i}{2} (\bp +\bq) \cdot \by 
 - \frac{1}{6} [(\bp \cdot \by)^2 + (\bq \cdot \by)^2] \right) \right. \nonumber \\
&& \left. + \frac{1}{12 \kf^2} (\bp^2 + \bq^2 + \bp \cdot \bq ) (3 \cos (z) + z \sin (z) - 3 \frac{\sin (z)}{z} ) Ê
 + \frac{1}{12} (\bp \cdot \by) (\bq \cdot \by) [  4 \frac{\sin (z)}{z} - 3 \cos (z)] \right] \nonumber
\eea
By comparison of these two expression to the expansions (\ref{N1_expansion}) and (\ref{N2_expansion}) we can directly read off the gradient coefficient functions of
Eq.(\ref{f_exp}) to be
\begin{align}
N^{1}_1 (y) &= -\frac{i}{2} j_0 (z) \quad \quad 
N^{1}_2 (y) =  \frac{1}{12 \kf^2} z \, j_1 (z) \quad \quad 
N^{1}_3 (y) =  - \frac{1}{6} j_0 (z) \nonumber \\
N^{2}_4 (y) &= \frac{\pi^2}{12 \kf^5} [ z^2 j_0 (z)- 3 z j_1 (z) ] \quad \quad 
N^{2}_5 (y) = \frac{\pi^2}{12 \kf^3} [ j_0 (z) + 3 z j_1 (z) ] \nonumber
\end{align}
where now we replaced $n_0  \rightarrow n(\br)$ and hence the Fermi wave vector $\kf = (3 \pi^2 n(\br))^{1/3}$ that appears in these
expressions is a spatially varying function. 
We further introduced the spherical Bessel functions
\be
j_0 (z) = \frac{\sin z}{z} \quad \quad 
j_1 (z) = \frac{\sin z - z \cos z}{z^2} . \nonumber
\ee
By inserting the gradient coefficient functions into Eq.(\ref{f_exp}) we find the following explicit gradient expansion for the 
one-particle density matrix
\bea
\gamma (\br',\br) &=&
\frac{\kf^3}{\pi^2} \frac{j_1 (z)}{z} + \frac{1}{2} j_0 (z) \by \cdot \nabla n(\br) - \frac{1}{12 \kf^2} z \, j_1 (z)  \nabla^2 n(\br) 
 + \frac{1}{6 \kf^2} z^2 j_0 (z) (\frac{\by}{y} \cdot \nabla)^2 n (\br) \nonumber \\
&& - \frac{\pi^2}{24 \kf^5} (z^2 j_0 (z) - 3 z j_1 (z)) \, (\nabla n(\br))^2  
 - \frac{\pi^2}{24 \kf^3}
( j_0 (z) + 3 z j_1 (z) ) \, (\by \cdot \nabla n(\br))^2 + \ldots
\label{1mat_final}
\eea
This expression is the main result of this section. After some manipulations we can rewrite it in an equivalent form as
\bea
\gamma (\br',\br) &=& \frac{1}{ \pi^2} \Big[  k_\F^3 \frac{j_1 (z)}{z} -  \frac{1}{24} \frac{\nabla^2 k_\F^2}{k_\F}
z j_1 (z) 
+ \frac{1}{12} \frac{1}{k_\F} \Big( \nabla \cdot (\nabla k_\F^2 \cdot \frac{\by}{y})\Big)  \cdot \frac{\by}{y} \,z^2 \, j_0 (z) 
+ \frac{1}{4} \nabla k_\F^2 \cdot \frac{\by}{y} \, z j_0 (z) \nonumber \\
&-& \frac{1}{96} \frac{(\nabla k_\F^2)^2}{k_\F^3} (j_0 (z) z^2 - z j_1 (z)) 
+ \frac{1}{32} \frac{1}{k_\F^3} \Big( \nabla k_\F^2 \cdot \frac{\by}{y} \Big)^2 (z^2 j_0 (z)-z^3 j_1 (z)) \Big]
\label{eq:gammas-exp}
\eea
This expression becomes identical in form to that derived by Gross and Dreizler \cite{GrossDreizler:81} using the Kirzhnits method after eliminating the
effective Fermi vector of their work in favor of the density (this requires inversion of Eq.(19) of reference \cite{GrossDreizler:81} ).
We close this section with a final remark on our result.
We note that the gradient expansion to finite order breaks the symmetry $\gamma (\br,\br')=\gamma (\br',\br)^*$.
However, the full gradient expansion as described by Eq.(\ref{eq:fexp2}) has this symmetry, which means that the symmetry is restored
by taking all higher order gradients into account whenever the series converges.  The symmetry violation has clearly consequences for the
quantities that are derived from it, such as the exchange hole, as it introduces ambiguity in defining such functions. To be in accordance with existing literature we will in the next section adopt 
the definition of the exchange hole that was used by Perdew \cite{Perdew:PRL85}.

\subsection{The exchange hole and energy}

We finally stress a few points on the properties of the exchange hole as derived from the gradient expansion and present an alternative derivation of
the gradient expansion for the exchange energy compared to that of Dreizler and Gross which has the advantage of avoiding the regularization of divergent integrals \cite{GrossDreizler:81}. In doing so we confirm a result obtained by Langreth and Perdew \cite{LangrethPerdew:PRB80}.
The exchange hole can be calculated directly from
the one-particle density matrix as
\be
\rho_{\x} (\br +\by  | \br) = -\frac{ |\gamma (\br + \by,\br)|^2}{2 n(\br)} . \nonumber
\ee
Inserting the expansion (\ref{1mat_final}) of for the one-particle density matrix and retaining terms to second order in the gradients yields the expression
\begin{align}
\rho_{\x} (\br +\by  | \br) &= - \frac{9}{2} n(\br) \frac{j_1^2 (z)} {z^2} - \frac{3}{2} \frac{j_0 (z) j_1 (z)}{z} (\by \cdot \nabla n(\br)) + 
\frac{1}{4 \kf^2} j_1 (z)^2 \nabla^2 n(\br) - \frac{1}{2 \kf^2} z j_0 (z) j_1 (z) (\frac{\by}{y} \cdot \nabla)^2 n (\br) \nonumber \\
& + \frac{\pi^2}{8 \kf^5} j_1 (z) (z j_0 (z) - 3 j_1 (z)) (\nabla n (\br))^2 + \frac{\pi^2}{8 \kf^3 } (\frac{j_1 (z) j_0 (z)}{z} + 3 j_1 (z)^2- 3 j_0 (z)^2) 
(\by \cdot \nabla n(\br))^2 + \ldots \label{xhole_grad}
\end{align}
As was pointed out by Perdew \cite{Perdew:PRL85} the second order gradient expansion of the exchange hole does not satisfy the negativity and sum rule 
constraints 
\be
\rho_\x (\br+ \by | \br) \leq 0 \quad \quad \int d\by \, \rho_\x (\br+ \by| \br)=-1   . \nonumber 
\ee
The violations of the negativity constraint is readily verified. However, the violation of the sum rule is not immediately obvious from Eq.(\ref{xhole_grad}).
The sum rule can, however, be conveniently analyzed in momentum space. To do this we define the Fourier transformed exchange hole to be
\be
\rho_\x (\bk | \br) = \int d\by \, \rho_\x (\br+ \by | \br)  e^{-i \bk \cdot \by}
\label{xhole_mom}
\ee
This function has the form
\begin{align}
\rho_\x (\bk | \br) = & \rho_\x^0 (\bk | \br) + \alpha_1 (k,n) \, \hat{\bk} \cdot \nabla n (\br) + \alpha_2 (k,n) \, \nabla^2 n(\br) + 
\alpha_3 (k,n) \, (\hat{\bk} \cdot \nabla n(\br))^2  \nonumber\\
& + \alpha_4 (k,n) (\nabla n(\br))^2  + \alpha_5 (k,n) \, (\hat{\bk} \cdot \nabla)^2 n(\br) + \ldots \label{momhole_exp}
\end{align}
where we defined the unit vector $\hat{\bk}=\bk /k$ and $k=|\bk|$.
The explicit form of the functions $\rho_\x^0$ and $\alpha_j$ follow directly by Fourier transforming Eq.(\ref{xhole_grad}) and are given in Appendix \ref{xhole_kspace}. The coefficient functions $\alpha_j$ are tempered distributions \cite{DuistermaatKolk} which have mathematically well-defined Fourier 
transforms. As follows directly from Eq.(\ref{xhole_mom}) the sum rule condition in momentum space translates to the requirement that $\rho_\x (\bk=0| \br)=-1$.
Since $\rho_\x^0 (\bk=0| \br)=-1$ (see Eq.(\ref{rho0_k}) ) this implies that the sum rule would be fullfilled whenever $\alpha_j (k=0,n)=0$. However, 
as is clear from Eqs.(\ref{alph1})-(\ref{alph5}) in the Appendix this is not satisfied for $\alpha_1$, whereas $\alpha_3$ and $\alpha_5$ even diverge for $k \rightarrow 0$  when interpreted locally as functions rather than distributions. The gradient expansion of the exchange hole does therefore indeed not satisfy the sum rule.
These divergencies, however, cancel when we calculate the system averaged exchange hole in momentum space which is given by the expression
\begin{align}
N \langle \rho_\x (\bk) \rangle  & = \int d\br \, n(\br) \rho_\x (\bk | \br) \nonumber \\
&= N \langle \rho^0_\x (\bk) \rangle + \hat{\bk} \cdot \int d\br \, n(\br) \alpha_1 (k,n(\br)) \nabla n(\br)
+ \sum_{i,j=1}^3  \int d\br \, n (\br) \beta_{ij} (k, n(\br) ) \partial_i n(\br) \partial_j n(\br)  .
\label{sa_khole}
\end{align}
\end{widetext}
where $N$ is the number of electrons.
In this expression $\langle \rho^0_\x (\bk) \rangle$ is the system average of $ \rho_\x^0 (\bk | \br) $ and in the last term under the integral sign we
defined the tensor
\be
\beta_{ij} (n,k) = \alpha_{\rm L} (n, k) \frac{k_i k_j}{k^2} + \alpha_{\rm T} (n, k) \Big( \delta_{ij} - \frac{k_i k_j}{k^2} \Big) . \nonumber
\ee
This tensor has a longitudinal part with coefficient $\alpha_{\rm L}$ and a transverse part with coefficient $\alpha_{\rm T}$ which
describe the contributions to the system-averaged hole of density gradients $\nabla n$ parallel and perpendicular to the momentum vector $\bk$.
These coefficients are calculated from the functions $\alpha_j$ as
\begin{align}
\alpha_{\rm T} = & \alpha_4 - \frac{1}{n} \frac{\partial (n \alpha_2)}{\partial n}  \nonumber \\
\alpha_{\rm L} = & \alpha_{\rm T} + \alpha_3 - \frac{1}{n} \frac{\partial (n \alpha_5)}{\partial n} , \nonumber 
\end{align}
as a short calculation on the basis of Eq.(\ref{momhole_exp}) will show.
From these equations and the explicit expressions for $\alpha_j$ given in Appendix \ref{xhole_kspace} we can readily calculate the explicit 
expressions for $\alpha_{\rm L,T}$. If we define
the dimensionless variable
$\bar{k}=k/(2 \kf)$ then we have
\begin{align}
\alpha_{\rm T} = & \frac{\pi^2}{32 n \kf^5} Z_{\rm T} (\bar{k}) \nonumber \\
\alpha_{\rm L} = & \frac{\pi^2}{32 n \kf^5} Z_{\rm L} (\bar{k}) , \nonumber
\end{align}
where the functions $Z_{\rm L,T}$ have the explicit form
\begin{align}
Z_{\rm T} (x) = & - 4 x \, \theta (1-x) + \frac{4}{3} \delta (x-1) \nonumber \\
Z_{\rm L} (x) = & - 4 x \, \theta (1-x) +  \delta (x-1) + \frac{1}{3} \delta' (x-1) \nonumber
\end{align}
These expressions are in accordance with the results of Langreth and Perdew (see Eq.(3.55) of reference \cite{LangrethPerdew:PRB80})
and this independent result therefore confirms the correctness of our alternative derivation. The interesting observation is now
that 
\be
\lim_{k \rightarrow 0} Z_{\rm L,T} (\bar{k}) = 0 \nonumber
\ee
and hence $\beta_{ij} (k,n) \rightarrow 0$ when $k \rightarrow 0$. This implies that the last term in Eq.(\ref{sa_khole})
does not contribute to the sum rule of the system averaged hole. The same conclusion can be derived for the second term in Eq.(\ref{sa_khole}).
Since $\alpha_1 (k=0,n)= i /(4 n \kf)$ (see Eq.(\ref{alph1})) is a local function of the density the integrand of the first integral 
in Eq.(\ref{sa_khole}) is a total derivative and vanishes (assuming either vanishing densities at infinity or periodic boundary conditions). 
We therefore find that
\be
\lim_{k \rightarrow 0} \langle \rho_\x (\bk) \rangle  = \lim_{k \rightarrow 0} \langle \rho_\x^0 (\bk) \rangle = -1 . \nonumber
\ee
This implies that system averaged exchange hole obtained from the second order gradient expansion does fulfill the sum rule, although the
exchange hole itself does not. We note, however, that this is only achieved by integrating over both positive and negative contributions.
When the positive contributions to the exchange hole are removed the sum rule is only recovered for a finite hole radius \cite{Perdew:PRL85,PerdewWang:PRB86}.
We finally note that with expression Eq.(\ref{sa_khole}) it is now a simple matter to calculate the exchange energy from
\be
E_\x [n] = \frac{N}{2} \int \frac{d \bk}{(2 \pi)^3} \langle \rho_{\x} (\bk) \rangle \frac{4 \pi}{|\bk|^2} . \nonumber
\ee
We insert Eq. (\ref{sa_khole}) and do the angular integrations first. It is therefore convenient to define the spherical and system averaged
hole in momentum space by
\be
 \llangle \rho_\x (k) \rrangle = \frac{1}{4 \pi} \int d \Omega_{\bk} \, \langle \rho_\x (\bk) \rangle \nonumber
 \ee
such that
\be
E_\x [n] = \frac{N}{\pi} \int_0^\infty dk \, \llangle \rho_\x (k) \rrangle .
\label{ex_spherical}
\ee
From Eq.(\ref{sa_hole}) we then find that
\begin{align}
 N & \llangle \rho_\x (k) \rrangle  =  N \llangle \rho_\x^0 (k) \rrangle \nonumber \\
& + \int d\br \, n(\br) [ \frac{1}{3} \alpha_{\rm L} (k,n)+ \frac{2}{3} \alpha_{\rm T} (k,n) ] (\nabla n)^2 \nonumber \\
& =  N \llangle \rho_\x^0 (k) \rrangle + \int d\br \, \frac{\pi^2}{32 \kf^5} Z_\x (\bar{k}) (\nabla n)^2 ,
\label{hole_spherical}
\end{align}
where we defined
\be
Z_\x (\bar{k}) = \frac{1}{3} Z_{\rm L} (\bar{k}) + \frac{2}{3} Z_{\rm T} (\bar{k}) .
\ee
The exchange energy is obtained by inserting the expression for the averaged hole of Eq.(\ref{hole_spherical})
into Eq.(\ref{ex_spherical}).
This then finally gives the expression
\begin{align}
E_\x [n] = &  -\frac{3}{4} \Big( \frac{3}{\pi} \Big)^{\frac{1}{3}} \int d\br \, n^{\frac{4}{3}} (\br) \nonumber \\
  & + 
\int d\br \, B_\x (n(\br)) (\nabla n (\br))^2
\label{ex_gradexp}
\end{align}
where
\begin{align}
B_\x (n) = & \frac{\pi}{16 \kf^4} \int_0^\infty d \bar{k} \, Z_\x (\bar{k}) \nonumber \\
= & - \frac{7}{432 \pi (3 \pi^2)^{1/3}} \frac{1}{n^{4/3}} \nonumber
\end{align}
and where to calculate the first term in Eq.(\ref{ex_gradexp}) we used Eq.(\ref{rho0_k}).
The coefficient $B_\x$ is the same gradient coefficient as obtained by Gross and Dreizler \cite{GrossDreizler:81} and 
earlier by Sham \cite{Sham:book}. However, the correct analytic exchange gradient coefficient is 
known \cite{Antoniewicz,Kleinman,EngelVosko,SvendsenBarth:IJQC95,LangrethVosko:AQC} to be a factor 10/7 
larger. The reason for this discrepancy is clearly described by Svendsen and von Barth \cite{SvendsenBarth:IJQC95} who showed that
the Coulomb interaction is too singular to allow for the interchange of the operations of integration and the
expansion in wave vectors. The problem does, for instance, not appear when Yukawa-screened Coulomb interactions are used \cite{SvendsenBarth:IJQC95}.
The conclusion is that there is nothing wrong with the Kirzhnits method, or the nonlinear response theory derivation of the gradient expansion
of the exchange hole presented here, but one has be aware that for Coulomb interactions the procedures of expanding correlation
functions in terms of density gradients followed by integrations involving Coulomb potentials may not yield the same result as directly expanding the integrated quantity in
terms of density gradients. 
For this reason the original GGA of Perdew and Wang \cite{PerdewWang:PRB86} based on the gradient expansion of the
exchange hole was later reparametrized \cite{Perdew:Physica,Perdew:book} to accomodate the correct gradient coefficient for the exchange energy.
 
\section{Conclusions and outlook}

We derived a general scheme based on nonlinear response theory 
to calculate the density gradient expansion of general correlation functions and showed that in order
to express the gradient coefficients in terms of the full density profile summations to infinite order must be carried out over
response functions of arbitrarily large order. A consistency condition was derived to do this. The formalism was used to derive the
gradient expansion of the one-particle density matrix and the exchange hole to second order in the gradients. We confirm the
derivation of Dreizler and Gross that used the Kirzhnits expansion method. 
We further analyzed the exchange hole in momentum space to derive that the system averaged hole to second order in the gradients
satisfies the sum rule and to derive the gradient expansion of the exchange energy without the need to regularize divergent integrals.

The scheme that we presented is very general and
can be applied to more general correlation functions. A natural application of the scheme would be to calculate the gradient expansion
of the correlation hole $\rho_\cc$. Regarding the gradient terms of the correlation hole little is known. We essentially only have some detailed
information on the long-range properties of the spherically and system averaged hole.  This information comes from the work of
Langreth and Perdew who calculated $\llangle \rho_\cc (k) \rrangle$ within the RPA. In our notation their results
(see also Appendix C of \cite{LangrethPerdew:PRB80}) can be summarized
as
\be
N \llangle \rho_\cc (k) \rrangle = N \llangle \rho_\cc^0 (k) \rrangle + \int d\br \, \frac{\pi^2}{16 \kf^4} z_\cc (n,k) (\nabla n)^2 \nonumber
\ee
where the function $z_\cc$ has been parametrized by Langreth and Mehl \cite{LangrethMehl:PRL81,LangrethMehl:PRB83}
as
\be
z_\cc (n,k) = \frac{4 \sqrt{3}}{k_{\rm s}} e^{- 2 \sqrt{3} \frac{k}{k_{\rm s}}}
\label{zcc}
\ee
where $k_{\rm s} =(4 \kf/\pi)^{1/2}$ is the Thomas-Fermi or screening wave vector.
We can transform to real space to obtain the following expression for the spherically and system averaged correlation hole
\begin{align}
N \llangle \rho_\cc (y) \rrangle & =  N \llangle \rho_\cc^0 (y) \rrangle \nonumber \\
& + \int d\br \frac{1}{96 \kf^4} \frac{k_{\rm s}^2}{(1+ (k_{\rm s} y)^2 /12)^2} (\nabla n)^2 
\label{chole_grad}
\end{align}
We see from Eq.(\ref{zcc}) that $z_\cc (n,k=0) \neq 0$ and as a consequence the sum rule $\llangle \rho_\cc (0) \rrangle=0$ for the correlation hole
is not satisfied. We know, however, that the RPA becomes accurate in the high density limit and since from Eq.(\ref{chole_grad}) the 
gradient coefficient of the correlation hole depends on $k_{\rm s}y$ we see that high densities are equivalent to large separations $y$. 
Similarly, low density corresponds to
short-distance behavior. However, RPA can not be trusted in this region and we have no precise knowledge on the small distance behavior
of the gradient coefficient of $\llangle \rho_\cc (y) \rrangle$. A model for the general short-range behavior was 
proposed \cite{Perdew:book,BPWbook,PBW:PRB96} on physical
arguments (it should not affect the Coulomb cusp of and the on-top value of the LDA hole) 
after which the real-space cutoff procedure was applied (see e.g. Fig. 4 in \cite{BPWbook} and Fig. 5 in \cite{PBW:PRB96})
to obtain a GGA for correlation. It would, however, be desirable to have a first principles approach to the calculation of the
short range properties of the gradient coefficient of the correlation hole. As follows from our derivation this requires
the knowledge of the functions (\ref{mathf1}) and (\ref{mathf2}), or at least their expansion to second order in wave vectors,
for the case that $F$ represents the pair correlation function or the correlation hole of the electron gas.
It may therefore be worthwhile to use our current scheme to explore these response functions beyond the RPA.
For short range correlation an approach based on diagrammatic summation of ladder diagrams suggests itself.
Of course, for the development of density functionals for general systems beyond the weakly inhomogeneous regime it
is not sufficient to use the straightforward gradient expansion \cite{Springer:PRB96}. However, general short-range features, such as the way the exchange-correlation hole
deforms close to the reference electron in the presence of a density gradient can be transferred to more general systems than the
weakly inhomogeneous ones.
Perhaps in combination with truly nonlocal functionals for the exchange-correlation hole \cite{GiesbertzSX} this can lead to the
development of more accurate density functionals. This will be a topic of our future research.

Finally, we would like to make some general remarks on the extension of our derivations to temperature- or time-dependent systems. 
In the case of temperature-dependent systems the expectation values of observables are traces over a grand canonical
ensemble. By Mermin's theorem \cite{Mermin} such expectation values are still functionals of the density 
and therefore the derivations in this work can be carried out unchanged.
In fact, the dependence on temperature is probably going to improve the convergence properties of the gradient series by smoothening of the
sharp Fermi functions which, for instance, were the cause of the oscillatory nature of the gradient coefficients of the 
one-particle density matrix. The situation of time-dependent systems is more delicate. Although the existence of a density-potential
mapping has been well-established \cite{RungeGross,vanLeeuwen,Ruggenthaler1,Ruggenthaler2} and hence a density functional theory 
can be formulated, the time-variable induces severe complications. As was shown by Vignale and Kohn \cite{VK1,VK2} the temporal and
spatial non-locality of time-dependent density functionals are intimately correlated. Any temporal non-locality (or frequency dependence
in the language of equilibrium response functions) induces long-range spatial properties that prevents the gradient expansion from existing.
A way out of this problem is to use new variables such as the current-density \cite{VK1,VK2} or the local density deformation density
in a Lagrangian frame \cite{Tokatly1,Tokatly2,Tokatly3} for which a local density approximation and a corresponding gradient expansion does exist.  
This has been exploited in Ref.\cite{Tao} to construct a GGA functional within time-dependent current density functional theory.
The study of correlation functions in the same fashion remains a challenge for the future.

\begin{acknowledgments}
I would like to thank the students in the course "Density Functional Theory" at the University of Jyv\"askyl\"a
for inspiration. I would further like to thank Esa R\"as\"anen for useful discussions and Klaas Giesbertz for checking a large part of the equations .
\end{acknowledgments}

\label{conclusions}

\appendix

\section{Calculation of $\gamma^1$ and $\gamma^2$}

\label{gammaderv}

In this Appendix we will derive the expression for the first and second derivative of the one-particle density matrix 
$\gamma$ with respect to the potential $v$. 
For a clearer interpretation and compactness of notation we will only put a comma between the variables representing the original
coordinates of the density matrix and the variables that appear as argument of the potential variations.
Direct differentiation of Eq.(\ref{onemat2}) using Eq.(\ref{phiderv}) then gives
\begin{align}
\gamma^1 (23,1) &= \frac{\delta \gamma (23)}{\delta v(1)} \nonumber \\
&= 2 \sum_{i,j}^{\infty} (f_j-f_i) \frac{ \varphi_j (1) \varphi_i^* (1) \varphi_i (2) \varphi_j^* (3)}
{\epsilon_j - \epsilon_i }  
\label{eq:gammas1}
\end{align}
where we used the short notation $j=\br_j$ and introduced the occupation numbers $f_j=1$ for an occupied state and 
$f_j=0$ for an unoccupied state. Inserting plane wave states appropriate for the electron gas we find that
\bea
\gamma^1 (\br_2 \br_3,\br_1) 
&=&  2  \int \frac{d \bq d\bp}{(2 \pi)^6} \frac{n_{\bp+\bq}-n_\bp}{\epsilon_{\bp+\bq}-\epsilon_\bp} 
\, e^{i \bp \cdot (\br_2-\br_3) + i \bq \cdot (\br_1 -\br_3)} . \nonumber
\eea
We therefore find that the function $\gamma^1$ has the simple form
\be
\gamma^{1} (\bp,\bq)= 2 \, \frac{n_{\bp+\bq}-n_\bp}{\epsilon_{\bp+\bq}-\epsilon_\bp} \nonumber
\ee
in Fourier space. This is precisely the integrand of Eq.(\ref{gamma_1}).
The function $\gamma^2$ can be calculated by straightforward differentiation
of Eq.(\ref{eq:gammas1}) with respect to the potential
\bea
\gamma^2 (23,14) =  \frac{\delta \gamma^1 (23,1)}{\delta v (4)} . \nonumber
\eea
To obtain the explicit form of this function we have to differentiate both the orbitals and the eigenvalues using Eqs. (\ref{epsderv}) and (\ref{phiderv}). Let us denote the term that arises from the differentiating the orbitals
by $X$ and the term that arises from differentiating the eigenvalues by $Y$. Then we find that
\be
\gamma^2 (23,14) = X (23,14) + Y (23,14) , \nonumber
\ee
where
\begin{align}
X (23,14)  & = 2  \sum_{i,j,k}^{\infty}  \varphi_i (2) \varphi_j^* (3) \nonumber \\
 & \Big[ \Phi (ijk) \varphi_k (1) \varphi_i^* (1)  \varphi_k^* (4)  \varphi_j (4)   \nonumber \\
 &  + \Phi (jik)  \varphi_j (1) \varphi_k^* (1)  \varphi_i^* (4)  \varphi_k (4) \Big] 
\label{X_1} \\
Y(23,14) &=  -2  \sum_{i,j}^{\infty}   \varphi_i (2) \varphi_j^* (3) \nonumber \\
 & \Phi (iji) ( | \varphi_j (4)|^2 - |\varphi_i (4)|^2 )   \varphi_j (1) \varphi_i^* (1) 
 \label{Y_1}
\end{align}
and where we further defined
\be
\Phi (ijk) =  \frac{1 }{\epsilon_k-\epsilon_j}  \left(     \frac{ f_k-f_i} {\epsilon_k - \epsilon_i } -  \frac{ f_j-f_i} {\epsilon_j - \epsilon_i }        \right) .
\label{Phi-fac}
\ee
The function $\Phi$ has the useful properties $\Phi (ijj)=0$ and $\Phi (ijk)=\Phi (ikj)$.
The function $\gamma^2 (23,14)= \gamma^2 (23,41)$ is symmetric in the indices $4$ and $1$ due to the fact that
the differentiations with respect to the potential commute. 
This is, however, not obvious from Eqs.(\ref{X_1}) and (\ref{Y_1}). We therefore want to rewrite the form of the function $\gamma^2$ 
in such a way that this symmetry is obvious. To do this we first note that since $\Phi (iji)=-\Phi (jij)$ the terms with $k=i$ and $k=j$ in Eq.(\ref{X_1}) sum up
to a function $Z$ of a similar form as $Y$ in Eq.(\ref{Y_1}), namely
\begin{align}
Z (23,14) & = -2  \sum_{i,j}^{\infty}  \varphi_i (2) \varphi_j^* (3) \nonumber \\
&  \Phi (iji) ( |\varphi_j (1)|^2 -
 |\varphi_i (1)|^2 )   \varphi_j (4) \varphi_i^* (4) .
\label{Z_1}
\end{align}
We can therefore write $\gamma^2 (23,14)$ as
\begin{align}
\gamma^2 (23,14) & = 2  \sum_{i j k, k \neq (i,j)}^{\infty}  \varphi_i (2) \varphi_j^* (3) \nonumber \\ 
& \Big[ \Phi (ijk) \varphi_k (1) \varphi_i^* (1)  \varphi_k^* (4)  \varphi_j (4) \nonumber \\
& + \Phi (jik)  \varphi_j (1) \varphi_k^* (1)  \varphi_i^* (4)  \varphi_k (4) \Big]  \nonumber \\
& + Y (23,14) + Z (23,14)
\label{gamma_s_2}
\end{align}
It is easily seen that the sum of $Y$ and $Z$ is symmetric under interchange of  $1$ and $4$. However,
this is not obvious in the first term of the equation above since at first sight $\Phi (jik)$ does not appear
to be equal to $\Phi (ijk)$ since
\be
\Phi (jik) =  \frac{1 }{\epsilon_k-\epsilon_i}  \left(     \frac{ f_k-f_j} {\epsilon_k - \epsilon_j } -  \frac{ f_i -f_j} {\epsilon_i - \epsilon_j }        \right) 
\label{Phi-fac_2}
\ee
which seems to be different from expression (\ref{Phi-fac}). However, this is just appearance. The reason is that the occupations
can only attain the values zero and one. For the case $f_i=f_j=1$ or $f_i=f_j=0$ we see directly that Eq.(\ref{Phi-fac}) and
(\ref{Phi-fac_2}) attain the same value. A little inspection shows that this is also true for cases $f_i=0, f_j=1$ and $f_i=1,f_j=0$.
We therefore find for $k \neq i,j$ that $\Phi (ijk) = \Phi (jik)$.
Therefore Eq.(\ref{gamma_s_2}) can be simplified to
\begin{align}
\gamma^2 (23,14) & = 2  \sum_{i j k, k \neq (i,j)}^{\infty}  \varphi_i (2) \varphi_j^* (3) \nonumber \\
& \Phi (ijk) \Big[  \varphi_k (1) \varphi_i^* (1)  \varphi_k^* (4)  \varphi_j (4) \nonumber \\
&+   \varphi_j (1) \varphi_k^* (1)  \varphi_i^* (4)  \varphi_k (4) \Big]  \nonumber \\
& + Y (23,14) + Z (23,14)
\label{gamma_s_22}
\end{align}
This expression is now explicitly symmetric in the indices $1$ and $4$. Let us now evaluate this expression for the
homogeneous electron gas. In the electron gas the one-particle states are plane waves 
\be
\varphi_{\bk} (\br ) =\frac{e^{i \bk \cdot \br}}{\sqrt{V}} \nonumber 
\ee
where $V$ is the volume of the system. Since $|\varphi_\bk|^2=1/V$ it follows from Eqs.(\ref{Y_1}) and (\ref{Z_1}) that the
terms $Y$ and $Z$ are identically zero and the function $\gamma (23,14)$ of Eq.(\ref{gamma_s_22}) therefore attains the form
\begin{align}
\gamma^2 (23,14)  & =  \frac{2}{V^3}  \sum_{\bk, \bp, \bq, \bq \neq (\bk,\bp)} 
\Phi (\bk,\bp, \bq)  e^{i ( \bk \cdot \br_2- \bp \cdot \br_3)} \nonumber \\
&\left[    e^{i (\bq-\bk) \cdot \br_1 + i (\bp-\bq) \cdot  \br_4 }   +  \, e^{i (\bp-\bq) \cdot \br_1 + i (\bq-\bk) \cdot  \br_4   } \right]
\label{gamma_s_3} \nonumber 
\end{align}
where we defined
\be
\Phi (\bk,\bp,\bq) =  \frac{1 }{\epsilon_\bq-\epsilon_\bp}  \left(     \frac{ n_\bq-n_\bk} {\epsilon_\bq - \epsilon_\bk } -  
\frac{ n_\bp- n_\bk} {\epsilon_\bp - \epsilon_\bk }        \right) . \nonumber
\ee
In this expression $n_\bp = \theta (\epsilon_{\F} - \epsilon_\bp)$ is the zero-temperature Fermi function and $\epsilon_{\F}=k_{\F}^2/2$ is the Fermi energy.
When we replace the sum by an integral the restriction $\bq \neq \bk,\bp$ is a region of measure zero and we can therefore freely
integrate over all wave vectors. After a few substitutions we obtain
\begin{align}
\gamma^2 (23,14) &=   \int \frac{d\bk d\bp d\bq}{(2 \pi)^9}  \gamma^{2} (\bk,\bp,\bq) \nonumber \\
& e^{i \bk \cdot (\br_2-\br_3)  + i \bp \cdot (\br_1-\br_3) + i \bq \cdot (\br_4-\br_3)}
\nonumber
\end{align}
where we defined
\be
 \gamma^{2} (\bk,\bp,\bq)  = 2 [ \Phi (\bk, \bk+\bp+\bq, \bk+\bp) +  \Phi (\bk, \bk+\bp+\bq, \bk+\bq) ] \nonumber
\ee
This gives the integrand of Eq.(\ref{gamma_2}).

\section{Expansion of $\gamma^{1}$ and $\gamma^{2}$}
\label{N1N2}

\subsection{Expansion of $\gamma^1$}

To determine $\gamma^1$ to second order in the wave vectors we have to expand
the function
\be
\gamma^{1} (\by, \bq) = 2 \int \frac{d \bp}{(2 \pi)^3} \frac{n_{\bp+\bq}-n_\bp}{\epsilon_{\bp+\bq}-\epsilon_\bp} 
\, e^{i \bp \cdot \by}
\label{gammas_1}
\ee
to second order in powers of $\bq$. This can be done by expanding the integrand in powers of $\bq$.
If we denote $\Delta= \epsilon_{\bp + \bq } - \epsilon_{\bp}
= \bp \cdot \bq + q^2/2$ with $q=|\bq|$, then we can expand the Fermi function $n_{\bp+\bq}$ in a distributional
Taylor series as
\be
n_{\bp + \bq} = n_{\bp} + \Delta \frac{d n }{d \epsilon}|_{\epsilon_{\bp}} + \frac{\Delta^2}{2} \frac{d^2 n }{d \epsilon^2}|_{\epsilon_{\bp}}
+ \frac{\Delta^3}{6} \frac{d^3 n }{d \epsilon^3}|_{\epsilon_{\bp}} + \ldots \nonumber
\ee
Since $n_\bp = \theta (\ef - \epsilon_{\bp})$ is the Heaviside function we have
\begin{align}
n_{\bp + \bq} & = n_{\bp} - \Delta \delta (\ef - \epsilon_{\bp}) + \frac{\Delta^2}{2} \delta' (\ef - \epsilon_{\bp}) \nonumber \\
& - \frac{\Delta^3}{6} \delta'' (\ef - \epsilon_{\bp}) + \ldots \nonumber
\end{align}
Inserting this into Eq.(\ref{gammas_1}) then gives
\begin{align}
\gamma^{1} (\by, \bq) &= -2 \int \frac{d \bp}{(2 \pi)^3} \, \delta (\ef - \epsilon_{\bp}) \, e^{i \bp \cdot \by} \nonumber \\
& + 
 \int \frac{d \bp}{(2 \pi)^3} \Delta \, \delta' (\ef - \epsilon_{\bp}) 
\, e^{i \bp \cdot \by} \nonumber \\
& - \frac{1}{3} \int \frac{d \bp}{(2 \pi)^3} \Delta^2 \, \delta'' (\ef - \epsilon_{\bp}) 
\, e^{i \bp \cdot \by} + \ldots
\label{gammas_11}
\end{align}
We see that in these integrals several derivatives of the delta function appear. We now use the general mathematical expression
\be
\delta^{(n)} (y(x)) = \left( \frac{1}{\frac{dy}{dx}} \frac{d}{dx} \right)^n \sum_i \frac{\delta (x-x_i)}{|\frac{dy}{dx} (x_i)|}
\nonumber
\ee
where the sum runs over all zeros of the function $y(x)$, i.e. $y(x_i)=0$. Using this equation in spherical coordinates
with $y(p)=(\kf^2-p^2)/2$ we find
\begin{align}
\delta (\ef - \epsilon_{\bp}) & = \frac{1}{\kf} \delta (p-\kf) \label{D0} \\
\delta' (\ef - \epsilon_{\bp}) & = - \frac{1}{p \kf} \delta' (p -\kf) \label{D1}\\
\delta^{''} (\ef - \epsilon_{\bp}) & = - \frac{\delta^{'}(p-\kf)}{\kf p^3} + \frac{\delta^{''} (p-\kf)}{\kf p^2} \label{D2}\\
\delta^{'''} (\ef-\epsilon_{\bp}) 
&= -\frac{1}{\kf} \Big[ \frac{1}{p^3} \delta^{'''} (p-\kf) - \frac{3}{p^4} \delta^{''} (p-\kf) \nonumber \\
&+  \frac{3}{p^5} \delta^{'} (p-\kf) \Big] \label{D3}
\end{align}
where in Eq.(\ref{D3}) we also evaluated the third derivative of the delta function as we will need it later
in the expansion of $\gamma^2$. 
With Eqs.(\ref{D0})-(\ref{D2}) we can readily evaluate the three integrals in Eq.(\ref{gammas_11}).
Let us denote these integrals by $A,B$ and $C$. Then we find for the first integral
\begin{align}
A &= -\frac{2}{\kf} \int \frac{d \bp}{\tpth} \delta (p - \kf)  \, e^{i \bp \cdot \by}  = -\frac{\kf}{\pi^2} \frac{\sin z}{z} \nonumber
\end{align}
where $z=\kf y$. The second integral gives 
\begin{align}
B & = \int \frac{d \bp}{\tpth} (\bp \cdot \bq + q^2/2) \delta' (\ef - \epsilon_{\bp}) 
\, e^{i \bp \cdot \by} \nonumber \\
& =
(q^2/2- i \bq \cdot \nabla_{\by}) \int \frac{d \bp}{\tpth} \delta' (\ef - \epsilon_{\bp}) \, e^{i \bp \cdot \by} \nonumber \\
 &= (q^2/2- i \bq \cdot \nabla_{\by}) \frac{1}{2 \pi^2 \kf} \cos (\kf y) \nonumber \\
&= \frac{q^2}{4 \pi^2 \kf} \cos z  + \frac{i \bq \cdot \by}{2 \pi^2 } \kf \frac{\sin z}{z} \nonumber
\end{align}
where in the second step we used Eq.(\ref{D1}).
It therefore remains to calculate the last integral
\be
C = -\frac{1}{3} \int \frac{d \bp}{\tpth} (\bp \cdot \bq + q^2/2)^2 \delta'' (\ef - \epsilon_{\bp}) 
\, e^{i \bp \cdot \by}  \nonumber
\ee
However, up to order $\bq^2$ it is suffices to calculate 
\begin{align}
C &= -\frac{1}{3} \int \frac{d \bp}{\tpth} (\bp \cdot \bq)^2 \delta'' (\ef - \epsilon_{\bp}) 
\, e^{i \bp \cdot \by}  \nonumber \\
&= \frac{1}{3} (\bq \cdot \nabla_{\by})^2 \int \frac{d \bp}{\tpth} \delta'' (\ef - \epsilon_{\bp}) 
\, e^{i \bp \cdot \by} \nonumber \\
&= - \frac{1}{3} (\bq \cdot \nabla_{\by})^2 \frac{1}{2 \pi^2 \kf^3} (\cos (z) + z \sin (z) ) \nonumber \\
&= - \frac{q^2}{6 \pi^2 \kf} \cos (z)  + \frac{(\bq \cdot \by)^2}{6 \pi^2 } \kf \frac{\sin (z)}{z} \nonumber
\end{align}
where in the second step we used Eq.(\ref{D2}).
Adding the results $\gamma^1=A+B+C$ gives the expression (\ref{gam1_result}).

\subsection{Expansion of $\gamma^2$}

Analogously to $\gamma^1$ we can expand $\gamma^2 (\by,\bp,\bq)$ of Eq.(\ref{gamma_2}) in powers of
$\bp$ and $\bq$. To reduce our effort we note that we can write
\be
\gamma^{2} (\by,\bp,\bq) = A (\by,\bp, \bq) + A (\by,\bq, \bp)  \nonumber
\ee
where
\be
 A (\by,\bp, \bq) = 2 \int \frac{d \mathbf{k}}{(2 \pi)^3}\, e^{i \bk \cdot \by} \Phi (\bk,\bk+\bp + \bq, \bk + \bp) \nonumber
\ee
We therefore only need to expand $A(\by,\bp,\bq)$ in powers of the wave vectors and symmetrize with respect to $\bp$ and $\bq$
afterwards. To do this we first need to expand the function 
\begin{align}
& \Phi (\bk,\bk+\bp + \bq, \bk + \bp) = \nonumber \\
&\frac{1}{\epsilon_{\bk +\bp}- \epsilon_{\bk+\bp+\bq}}
\left[ \frac{n_{\bk+\bp} - n_{\bk}}
{\epsilon_{\bk +\bp} - \epsilon_{\bk}}Ê- \frac{n_{\bk+\bp+\bq} - n_{\bk}}{\epsilon_{\bk +\bp + \bq} - \epsilon_{\bk}} \right] . \nonumber
\end{align}
If we denote
\begin{align}
\Delta_1 &= \epsilon_{\bk +\bp} - \epsilon_{\bk} = \frac{p^2}{2} + \bp \cdot \bk \nonumber \\
\Delta_2 &= \epsilon_{\bk +\bp+\bq} - \epsilon_{\bk} = \frac{(\bp+\bq)^2}{2} + (\bp+\bq) \cdot \bk , \nonumber
\end{align}
we find by expanding the Fermi functions in a distributional Taylor series that
\begin{align}
& \Phi (\bk,\bk+\bp + \bq, \bk + \bp)  = 
\frac{1}{2} \delta^{'} (\ef - \epsilon_{\bk})  \nonumber \\
& - \frac{1}{6} (\Delta_1+ \Delta_2) \delta^{''} (\ef - \epsilon_{\bk}) \nonumber \\
& + \frac{1}{24} (\Delta_1^2 + \Delta_1 \Delta_2 + \Delta_2^2) \delta^{'''} (\ef - \epsilon_{\bk}) + \ldots \nonumber
\end{align}
With this expression we find that the function $A(\by,\bp,\bq)$ can be calculated as
\begin{align}
& A (\by,\bp, \bq) \nonumber \\
& =\int \frac{d \bk}{(2 \pi)^3}\, \delta^{'} (\ef - \epsilon_{\bk}) \, e^{i \bk \cdot \by} \nonumber \\
&- \frac{1}{3}   \int \frac{d\bk}{(2 \pi)^3}\, \delta^{''} (\ef - \epsilon_{\bk}) (\Delta_1+ \Delta_2) \, e^{i \bk \cdot \by} \nonumber \\
&+ \frac{1}{12} \int \frac{d \bk}{(2 \pi)^3}\, \delta^{'''} (\ef - \epsilon_{\bk}) (\Delta_1^2+\Delta_1 \Delta_2+ \Delta_2^2) \, e^{i \bk \cdot \by} 
\nonumber \\
&+ \ldots \nonumber
\end{align}
The three integrals appearing on the left hand side of this expression are sufficient to extract the powers to second order in $\bp$ and $\bq$. Let us call these three integrals
$A_1$, $A_2$ and $A_3$ in order of appearance. The first integral gives
\begin{align}
A_1 &= \int \frac{d\bk}{(2 \pi)^3}\, \delta^{'} (\ef - \epsilon_{\bk}) \, e^{i \bk \cdot \by} = \frac{\cos z}{2 \pi^2 \kf} . \nonumber
\end{align}
The second integral is given by
\begin{widetext}
\begin{align}
A_2 &= - \frac{1}{3}\int \frac{d \mathbf{k}}{(2 \pi)^3}\, \delta^{''} (\ef - \epsilon_{\bk})
\left[ \frac{p^2+ (\bp+\bq)^2}{2} + (2 \bp + \bq) \cdot \bk \right]  \, e^{i \bk \cdot \by} \nonumber \\
&= \left[Ê-\frac{p^2+ (\bp+\bq)^2}{6} + \frac{i}{3} (2 \bp + \bq) \cdot \nabla_{\by} \right] 
\int \frac{d \mathbf{k}}{(2 \pi)^3}\, \delta^{''} (\ef - \epsilon_{\bk}) \, e^{i \bk \cdot \by} \nonumber \\
&= \left[Ê-\frac{p^2+ (\bp+\bq)^2}{6} + \frac{i}{3} (2 \bp + \bq) \cdot \nabla_{\by} \right] \frac{\cos (z) + z \sin (z) }{-2 \pi^2 \kf^3} \nonumber \\
&= \frac{\cos (z) + z \sin (z)}{ 12 \pi^2 \kf^3 } ( p^2+ (\bp+\bq)^2 )- \frac{i}{ 6 \pi^2 \kf } (2 \bp + \bq) \cdot \by \, \cos (z) \nonumber
\end{align}
where we used Eq.(\ref{D2}) in the second step. Finally the third integral has the explicit form
\begin{align}
A_3 &=  \frac{1}{12} \int \frac{d \mathbf{k}}{(2 \pi)^3}\, \Big[ (\frac{p^2}{2} + \bp \cdot \bk)^2 + 
(\frac{p^2}{2} + \bp \cdot \bk) (\frac{(\bp + \bq)^2}{2} + (\bp+\bq) \cdot \bk)   \nonumber \\
&  + (\frac{(\bp + \bq)^2}{2} + (\bp+\bq) \cdot \bk)^2 \Big] \delta^{'''} (\ef - \epsilon_{\bk}) \, e^{i \bk \cdot \by} \nonumber
\end{align}
However, we only need the terms to second order in $\bp$ and $\bq$. Therefore, it is sufficient to calculate
\bea
A_3 &=&  \frac{1}{12} \int \frac{d \mathbf{k}}{(2 \pi)^3}\, \left[ (\bp \cdot \bk)^2 + (\bp \cdot \bk) ((\bp+\bq) \cdot \bk)
+ ((\bp+\bq) \cdot \bk)^2 \right] \delta^{'''} (\ef - \epsilon_{\bk}) \, e^{i \bk \cdot \by} \nonumber \\
&=& -\frac{1}{12} \left[ (\bp \cdot \nabla_{\by})^2 + (\bp \cdot \nabla_{\by}) ((\bp+\bq) \cdot \nabla_{\by})
+ ((\bp+\bq) \cdot \nabla_{\by})^2 \right] \int \frac{d k}{(2 \pi)^3}\, \delta^{'''} (\ef - \epsilon_{\bk}) \, e^{i \bk \cdot \by} \nonumber \\
&=& -\frac{1}{12} \left[ (\bp \cdot \nabla_{\by})^2 + (\bp \cdot \nabla_{\by}) ((\bp+\bq) \cdot \nabla_{\by})
+ ((\bp+\bq) \cdot \nabla_{\by})^2 \right] 
 \left( \frac{3 \cos (z) - z^2 \cos (z) + 3 z \sin (z)}{2 \pi^2 \kf^5} \right) \nonumber
\eea
where to evaluate the integral over the delta function we used Eq.(\ref{D3}). To perform the derivatives in the last term we use
that for an arbitrary function $f(z)$
\be
(\bp \cdot \nabla_{\by})(\bq \cdot \nabla_{\by}) f(z) 
= (\bp \cdot \by)(\bq \cdot \by) \frac{\kf^2}{y^2} \left[ \frac{d^2 f}{dz^2} - \frac{1}{z} \frac{df}{dz}\right]
+ (\bp \cdot \bq) \frac{\kf}{y} \frac{df}{dz} \nonumber
\ee
With this expression we find that
\bea
A_3 &=&  - \frac{1}{24 \pi^2 \kf^3} (p^2 + \bp \cdot (\bp+ \bq) + (\bp+ \bq)^2 ) (\cos (z) + z \sin (z)) \nonumber \\
&& - \frac{1}{24 \pi^2 \kf} [  (\bp \cdot \by)^2 + (\bp \cdot \by) ((\bp+\bq) \cdot \by)
+ ((\bp+\bq) \cdot \by)^2   ] \cos (z) \nonumber
\eea
Collecting our results $A=A_1+A_2+A_3$ we therefore obtain the following expression for $\gamma^{2}(\by,\bp,\bq)$
\bea
\gamma^{2}(\by,\bp,\bq) &=& A (\by,\bp,\bq)  +  A (\by,\bq,\bp ) \nonumber \\
&=& \frac{1}{\pi^2 \kf} 
\left( \cos (z) - \frac{i}{2} (\bp + \bq) \cdot \by \cos(z) 
+ \frac{1}{12 \kf^2} (\bp^2+\bq^2 + \bp \cdot \bq) (\cos (z) + z \sin(z)) \right. \nonumber \\
&& \left. - \frac{1}{24} [  (\bp \cdot \by)^2 + (\bq \cdot \by)^2 + 3 ((\bp+ \bq) \cdot \by)^2 Ê] \cos (z) \right] \nonumber
\eea
This is equal to expression (\ref{gam2_result}).

\section{Exchange hole in momentum space}
\label{xhole_kspace}

Here we present the explicit expressions for the coefficient functions $\alpha_j$ of Eq.(\ref{momhole_exp}) 
which are calculated by Fourier transforming Eq.(\ref{xhole_grad}).
If we define the dimensionless quantity $\bar{k} = |\bk|/ (2 \kf (\br))$ then $\rho_\bx^0$ is given by
\be
\rho_\bx^0 (\bk | \br) = (-1 + \frac{3}{2} \bar{k} - \frac{1}{2} \bar{k}^3) \theta (1 - \bar{k})
\label{rho0_k}
\ee
which is simply the Fourier transform of the LDA exchange hole.
We see that $\rho_\bx (0 |\br)=-1$ and therefore the LDA hole satisfies the sum rule.
The functions $\alpha_j$ are given by
\begin{align}
\alpha_1 =&  \frac{i}{4 n \kf} \theta (1-\bar{k}) \label{alph1}\\
\alpha_2 =& - \frac{\bar{k}}{12 n \kf^2} \theta (1-\bar{k}) \label{alph2} \\
\alpha_3 =& - \frac{1}{72 n^2 \kf^2} \Big( \frac{\theta (1-\bar{k})}{\bar{k}} + \frac{1}{4} \delta (\bar{k}-1) 
+ \frac{3}{4} \delta'(\bar{k}-1) \Big)  \label{alph3} \\
\alpha_4 =& \frac{1}{72 n^2 \kf^2} \Big( 3 \bar{k} \, \theta (1-\bar{k}) - \delta (\bar{k}-1)  \Big) \label{alph4} \\
\alpha_5 =& \frac{1}{24 n \kf^2} \Big(   \frac{\theta (1-\bar{k})}{\bar{k}} +  \delta (\bar{k}-1) \Big) \label{alph5}
\end{align}

\end{widetext}

\end{document}